\documentclass[aps,showpacs,prl]{revtex4}
\usepackage{graphics}

\begin{document}

\count255=\time\divide\count255 by 60
\xdef\hourmin{\number\count255}
  \multiply\count255 by-60\advance\count255 by\time
 \xdef\hourmin{\hourmin:\ifnum\count255<10
0\fi\the\count255}

\title{\boldmath{The Narrow $\Theta(1543)$--A QCD
Dilemma:
Tube or Not Tube? }}

\author{Aharon Casher}
\email{ronyc@post.tau.ac.il}

\author{Shmuel Nussinov}
\email{nussinov@ccsg.tau.ac.il}

\affiliation{School of Physics and Astronomy, Tel Aviv
University, Tel Aviv, Israel\\
and\\
Department of Physics and Astronomy, University of
South Carolina, Columbia SC 29208}

\date{September 17, 2003}

\begin{abstract}

We argue that a width of less than MeV of the new
$\Theta$ resonance
is inconsistent with the observed ratio of resonance
and background events
in the various photon initiated experiments if the
latter can be described by $K K^*$, etc., exchange.
An evaluation of the Feynman diagrams
which were believed to be relevant is presented and
supports the general
claim in the one case where a cross section has been
given by the experimental group.

More detailed arguments
based on the flux tube model explaining the  narrow
widths and the apparent conflict
with the production rates are presented. We predict
narrow Tetra-quarks
at mass $\sim$ O(1-1.2 GeV) which the analysis of LEAR
may have missed.

\end{abstract}

\maketitle

\noindent {\large \bf I. General Phenomenological
Considerations}

Enhancements in $K^+n$ and $K^0p$ invariant mass at
$M(KN) \sim$ 1540  MeV
seen in several experiments \cite{TN,ST,JB,VB} using
different
reactions
with a range of incident energies, different
detectors with different acceptances and cuts suggest
a new, exotic ``Penta-quark". Such a low-lying, narrow
state,
\begin{equation}
   \Gamma(\Theta) < {\rm Exp \; resolution} \sim 20 \;
{\rm MeV} \, ,
\label{GammaThetaEq1}
\end{equation}
with $J^P=1/2^+$ has been predicted \cite{DPP} using
Skyrmion large
N approach \cite{T.Cohen} and ipso facto explained in
simpler ways \cite{N,JW,KL}.
It may mark the beginning of  a new ``era of exotics"
in hadron
physics.

Two puzzles emerge in connection with the width
$\Gamma(\Theta)$.
The eventual resolution of both may be a triumph of
(non-perturbative!) QCD.

The first puzzle is:

P(1): The absence of any indications for the new $KN$
resonance $K^+$
deuteron scattering
implies an anomalously low $\Theta$ decay width
\cite{N,ASW}:
\begin{equation}
      \Gamma(\Theta(1543)) < 0(MeV)
\label{GammaThetaEq2}
\end{equation}
The second is  the following: Within $K$ exchange
models, significantly
higher values of
$\Gamma(\Theta)$ are inferred from production rates in
photonic reactions.  We elaborate on this point next.

Let us assume that for the
$\theta$ production experiments, all done at medium
energies,
the hadronic (rather than perturbative
QCD) description is more appropriate.
The $\theta$ then forms via $KN$ or $K^*N$
intermediate
processes with the $K$, say, being relatively close to
its mass shell.
If $K$ exchange dominates, we can estimate
$\Gamma(\Theta)$ from the
production cross sections.  For  meaningful
comparisons the Feynman diagram calculations should be
done in
parallel with MC simulations including acceptance
and signal improvement cuts. This is beyond the scope
of the present
paper and the capabilities of the authors.

We can nonetheless estimate  $\Gamma(\Theta)$
independently
of these complications. We assume that the
$K$-exchange model holds
equally well for $KN$ invariant mass in the  ``true"
resonance region:
\[
    m(KN)= m(\Theta) \pm \Gamma(\Theta)
\]
and within the broader region of effective width
$\Gamma$(obs) where enhancements
in the  experimental $KN$ invariant mass,
$m(KN)$, distributions were observed.

The pion exchange model for the reaction $\pi$ +
proton
$\rightarrow \pi \pi$ + Nucleon
applies both off and on $\pi-\pi$ resonances and is
used
to map  $\pi-\pi$ scattering. When
extrapolated to the on-shell pion limit the reaction
rate at a given invariant mass $m(\pi-\pi)$ is
proportional to the $\pi-\pi$ cross section at this
CMS energy.
All we need is that the $K$ exchange share these
qualitative
features.

The number of events in the ``enhancement" (lying
above a
smooth curve interpolating between the regions to the
right and
left), $N(R)$, is identified with the number of
$\Theta$s, and the
remaining $N(B)$ events in the same region under this
curve represent
the non-resonant slowly varying background.

Integrating the Breit-Wigner  distribution of the
resonance with the
``true" narrow width $\Gamma(\Theta)$ the expected
$N(R)$ is:
\begin{equation}
      N(R) = F \cdot [\Gamma(\Theta)/2] \pi \sigma(R)
\, ,
\label{NReq3}
\end{equation}
where $\sigma(R)$, the peak resonance cross section,
is $2 \pi/k^2 \sim 33$ mb.
Likewise, the number of background events under the
peak expected in the same $K$ exchange model should
be:
\begin{equation}
           N(B)= F \cdot \Gamma{\rm (obs)} \cdot
\sigma(B)
\label{NBeq4}
\end{equation}
$\sigma(B)$, the off resonance $K(+)$ neutron total
cross section, is $\sim$ 14 mb at these energies
\cite{PDG}.
The common factor $F$ representing
dynamical/kinematical
aspects of the  computed Feynman diagram  and/or cuts
applied to
 events in the enhancement region
cancels in  the ratio of the last two equations and
\begin{equation}
  \Gamma(\Theta) = [N(R)/N(B)] \cdot
(\sigma(B)/\sigma(R)) \cdot \pi/2
        \cdot [\Gamma{\rm (obs)}] \sim [N(R)/N(B)]
\cdot 1/2
[\Gamma{\rm (obs)}] \; .
\label{NRNBeq5}
\end{equation}
Equation (\ref{NRNBeq5}) constitutes the second
puzzle:

P(2): Even for a minimal $\Gamma{\rm (obs)} \sim$ 20
MeV effective width  of the $\sim$ 4  bins enhancement
region,  the observed $N(R)/N(B)$ which exceeds .5 in
all the experiments, implies
\begin{equation}
  \Gamma(\Theta) \sim 5-15 \; MeV \;\;\; {\rm
(estimate \;  based \; on \;K \; exchange \; model)}
\label{GammaTheta5MeVeq6}
\end{equation}
conflicting with the upper bound of Eq.
(\ref{GammaThetaEq2}) above.
Would the apparent difficulty be evaded if $\theta$
production is not dominated by exchanging $K(490)$,
but
rather by the vector $K*(890)$ exchange or the tensor
$K(1420)$
exchange, etc.?
Even this in itself  is insufficient and a large O(10)
double ratio of resonant and non-resonant $K(*)N
\rightarrow KN$ and $KN \rightarrow KN$ cross sections
is required.

\noindent {\large \bf \boldmath{II. Calculations of
$\Theta$ Production
Rates Via $K$ Exchange in Photon-Nucleon and
Photon-Deuteron
Collisions}}

For completeness we present the cross section for
$\Theta$ production
in $\gamma-p/n$ collisions within a $K$ exchange model
(with possible rescattering on the remaining $n/p$ for
deuteron targets) in:
\begin{itemize}
\item (a) $\gamma + p \rightarrow \Theta + K(S)$
[Saphir]

\item (a') $\gamma + d \rightarrow \Theta +K^-+p$
[Spring 8][Jeff Lab]

\item (b) $\gamma + p \rightarrow \Theta +
K^- \pi^+$ with the final kaon
and pion in the $K^*(890)$ resonance.[Jeff Lab]
\end{itemize}

In the $\Theta$ discovery  in
(a') by the Spring 8
collaboration, the final $K^-$ could be very forward
and the undetected final proton have very low energy
as $K$ exchange with a spectator proton implies.
The same holds for the forward-going $K(S)$ in Saphir
but not for the class
detector.  Its limited forward coverage required
measuring the final protons which is possible only if
$k(p{\rm (final)}) > \sim 0.35$  GeV.  The rate
observed is
then suppressed by the small
probability of having such momentum in the
deuteron. Also the
diagram where the final $K^+/K^-$ re-scatters on $n/p$
to form $\theta$ / Kick the $K^-$ and $p$, is
suppressed
by the (related!)
extended configuration space wave function of the
deuteron.

We next sketch the computations starting
with the $K$-exchange ``tree" diagram in Fig 1.

\newpage
\clearpage


\begin{figure}[thb!]
\begin{center}
\vspace*{1.75in}
\mbox{
\includegraphics[8.75in,8.75in][0in,0in]{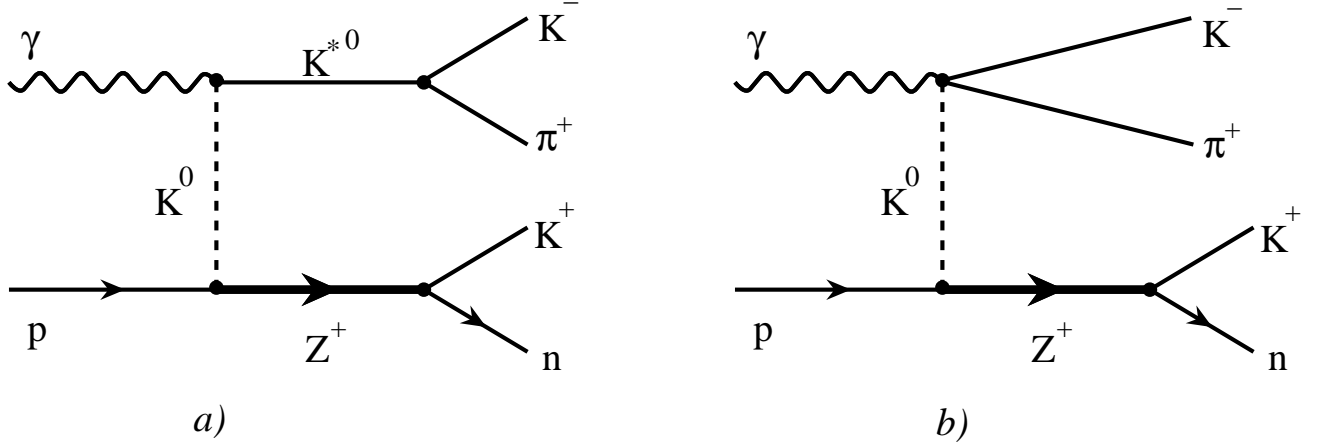}}
\vspace*{.75in}
\caption{{\footnotesize  The one-Kaon exchange diagram for
$Theta$ (denoted here as Z) production, for the particular
case when the latter is produced of a
proton and in association with a neutral $K^*(890)$,
namely, process (b) listed at the beginning of Sec II. The
variant of this diagram with the produced $K^*$ replaced by
a neutral Kaon is relevant for process (a). Finally a $K^+$
is exchanged in $\gamma+n \rightarrow K^-\Theta^+$.  All
diagrams share the same $NK\Theta$ lower vertex but differ in the
magnitude and/or form of coupling in the upper vertex. Note that
the suffixes (a) and (b) in the figure are {\it not}
associated with (a) and (b) in the text.  Figure 1(b) is drawn
to illustrate an off resonance (here $K^0$ of scattering).}}
\bigskip
\label{gammap}
\end{center}
\end{figure}

The coupling $g = g(\Theta N K)$, the analog of
$g[NN \pi] \sim [4 \pi \cdot 14]^{(1/2)} \pi$
nucleon $\gamma(5)$ coupling in the
lower vertex of Fig. 1(a) and 1(b) is fixed by
$\Gamma(\Theta)$.
The $\gamma k^0 K^0$ and $\gamma K^+ K^-$ coupling
in the upper vertex are $e/3$ and $e$ for  reactions
(a) and
(a'), respectively, and the $g^* K^* K^- \gamma$
coupling for reaction (b) is fixed by
$\Gamma(K^*) \rightarrow (K+\gamma)$ = 0.115 MeV.
This yields:
\begin{eqnarray}
   d(\sigma)/d(t)\{\gamma + N \rightarrow \Theta + K
\} & = &
    F \alpha \{(1/2)\Gamma(\Theta)/0.0226 \} \nonumber
\\
    (1/2)  \{2[t+(m[\theta]-m[N])^2]\} & \cdot &
     \{p(f)^2[2-(t-t[-]])/2p(f)p(i)\}/(t+m[K]^2)^2
\label{dsigmaNeq7}
\end{eqnarray}
\begin{eqnarray}
  d(\sigma)/d(t)\{\gamma + p \rightarrow
\Theta+K^*(890)\} & = &
   F \cdot (6.44 \cdot 10^{-3})(1/2)
(\Gamma(\theta)/0.0226)
\nonumber\\
   (2/3)(1/2)\{2[t+(m[\Theta]-m[N])^2]\} & \cdot &
   \{(t+m(K^*)^2)^2\}/(t+m[K]^2)^2
\label{dsigmapEq8}
\end{eqnarray}
$F = \pi/[(m(N)^2) \cdot (E(\gamma))]^2$ is the
flux factor. The first and second \{ \} brackets were
generated by $N-\Theta$ spinor and photon/$K^*$ polarization
sums. $p(f) p(i)$ in Eq. (\ref{dsigmaNeq7}) are the
final/initial center mass momenta of the $K$/photon.
The [square] of the Kaon propagator appears in the denominators,
and the momentum transfer $t$, the virtual kaon squared
momentum, varies between $t(-)$ and $t(+)$.
Since $\Theta$ decays equally to $K^+ n$ and $K^0 p$
we use 1/2 $\Gamma$ in
determining $g(\Theta \, N \, K)^2$, and  (2/3)/(1/2)
are branchings for $K^* \rightarrow K^+ \pi^-$
and  $\Theta \rightarrow K^+n$.

Comparing the integral of (\ref{dsigmaNeq7})
between $t(-)$ and $t(+)$
with the 300 $nb$ cross section quoted in Saphir
we obtain
\begin{equation}
        \Gamma > 30\; {\rm MeV} \;\;\;\;\;\;
{\rm (Saphir \; exp \; and \; the \; kaon \; exchange
\; model)}
\label{SaphirExpEq9}
\end{equation}
(An  inequality arises since form factors  suppressing
the vertices with off-off shell kaons have been
omitted.)

Equation (\ref{dsigmapEq8}) applies to $\gamma+p
\rightarrow \Theta+K^*$.

We next consider the one-loop diagrams like Fig. 2 for
$\Theta$ production off deuterons. These complex
diagrams with ``anomalous thresholds" can be estimated since the
deuterons' size $R$ ($\sim$ 2 Fermis) exceeds all
other distances in the problem. The kaon traveling a large
distance from production
on the proton/neutron to rescattering on the remaining
neutron/proton is effectively on shell. The process
$\gamma + d \rightarrow \bar{K}\bar{K} pn$ then factorizes
into two parts as explained next for $K$-neutron re-scattering.


\begin{figure}[thb!]
\begin{center}
\vspace*{1.75in}
\mbox{
\includegraphics[3.8in,3.8in][4in,4in]{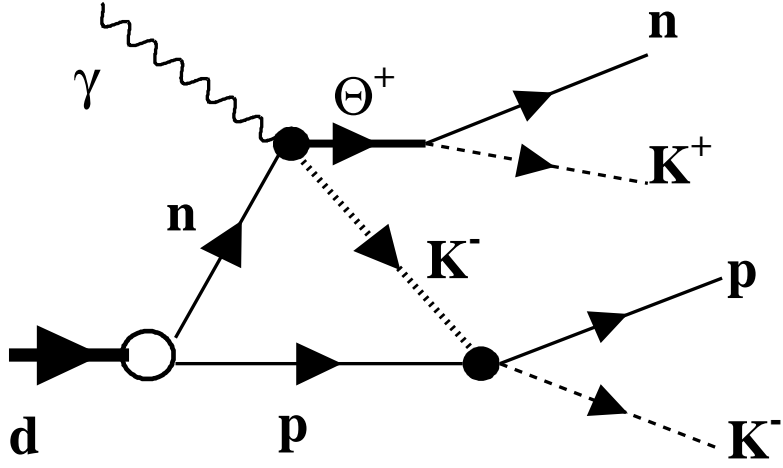}}
\vspace*{.75in}
\caption{{\footnotesize $\Theta$ production on the deuteron with
a re-scattering on the  nucleon. The diagram shown refers
to re-scattering of $K^-$ on the proton
(lower right black dot) after $\Theta^+ K^-$ have
been produced off the neutron.  The upper left black dot
indicating the latter process could be
dominated  by $K$ exchange diagram of the type shown
in Fig 1, or involve more  general local interactions which
cannot be represented by
$K,\; K^*(890), \; K(1420)$, etc., exchanges. A similar
diagram with $K^- \rightarrow K^+$ and $n,p\rightarrow p,n$
describes the process where the primary $\gamma p$
collision generates out-going $pK^+K^-$ with $E(K^+)$, the laboratory
energy of the $K^+$ constrained to be at resonance so that
the $\Theta$ is produced in the $K^+$ re-scattering off the
``spectator" neutron. It is this last process
that is discussed first in Sec. II above.}}
\bigskip
\label{diagram}
\end{center}
\end{figure}

Assume that we first have the process
$\gamma + p \rightarrow K^+K^-p$
(rather than the $\gamma N$ interaction depicted in
Fig 2. This then serves as a source of $K^+$'s
with energies $E(K^+)$ emanating from $r(p)$ where the
struck proton was.  Alongside the $K^+$ of interest
emerge also the $K^-$ and proton at momenta $P(p)$ and
$P(K^-)$ which are
unaffected in the next scattering.
Next, the $K^+$ scatters on the neutron which, in the
first scattering, was just a spectator located at
$r(n)$
at a distance $R = \mid r(p)-r(n) \mid$.

In the second scattering $\Theta$ manifests via the
enhanced $K^+n$
resonant scattering cross section at invariant masses:
$m(KN) = m(\Theta) \pm \Gamma(\theta)/2$.
The two processes are then compounded classically by
multiplying probabilities yielding:
\begin{eqnarray}
   d(\sigma\{\gamma +d \rightarrow K^- p \Theta]\} & =&
   \{< 1/[4 \pi R^2]>\} \nonumber\\
     \cdot d\{\sigma[\gamma + p \rightarrow K^-PK^+]\}/d(E(K^+)
     \cdot \{\Gamma(\theta) \cdot \pi/2\}
     \cdot \{m(N)/m(\Theta)\} \cdot \sigma[KN]{\rm (Res)}
\label{dsigmaKNResEq10} 
\end{eqnarray}

The $d(\sigma)$ on the right-hand side
refers to a differential (or partially integrated)
cross section with respect to the momenta of the $K^-$
and proton
which do not participate in the second collision.
In the case of $K-n$ re-scattering considered here
first, we fix the
the  energy $E(K^+)$ of the almost on-shell Kaon
to correspond to the $\Theta$ resonance in
the collision with the almost stationary neutron.
Thus the differential cross
section with respect to $E(K^+)$ is evaluated at the
resonant energy.
The BW integral then yields
$(\pi/2) \Gamma \sigma[KN]$(res) with the resonant
(Peak)
cross section $\sim$ 32 mb, $m(N)/m(\Theta)$ is the
Jacobian of the
transformation from
invariant mass to lab energy in the $K-N$ collision,
 and $\sigma(KN)/(4 \pi R^2)$ is the probability that
the
$K^+$ emitted from $r(p)$ will scatter at $r(n)$.  We
 use the expectation value $< >$ in the (isotropic)
deuteron ground state.

To evaluate the $\Theta$ production
cross section we need to input information on the
differential/partially integrated (apart from the
$E(K^+)$
dependence) $\gamma + p \rightarrow K^+K^-p$
cross sections, available from other experiments.

For $K^-$ re-scattering--which is, in fact, the one
depicted
in Fig. 2--the first process is
$\gamma + n \rightarrow \Theta K^-$ with a
spectator proton.  In the $K$ exchange model it
yields mostly forward-going $K^-$ and slow
final protons. The second $K^-p$
scattering  ``Kicks" the $K^-$ away from the forward
direction and
augment the protons' kinetic energy: $E(p)-m[N] \sim
t/2m[N]$,
making both visible in the Class detector. Using
similar arguments
as in the previous case we find:
\begin{equation}
  d(\sigma\{\gamma + d \rightarrow K^- p \Theta\}) =
  \{<1/R^2>\} \cdot d\{\sigma[\gamma + n \rightarrow
K^-
\Theta]\}/d(E(K^-)
  \cdot \sigma\{K^-p\}(E(K^-))
\label{dsigmaKEeq11}  
\end{equation}

The coupling of a photon to a
charged Kaon is $\sim$ an order of magnitude stronger
than that to the
neutral Kaon). Hence $\Gamma(\Theta)$ emerging from an
eventual
cross section supplied by the Class collaboration may
be somewhat smaller
than  the width implied by the Saphir cross sections.
The general
considerations of Sec. I
suggest, however, that also here the required $\Gamma$
may be unacceptably high

\noindent {\large \bf III. Color Suppression of Tetra-
and Penta-Quarks
in the Chromoelectric Flux Tube Model}

$\Gamma(\Theta) <$  0(MeV) is low for
$m(\Theta)-[m(K)+m(N)] \sim 110 \; MeV$. Our
discussion
above reinforces this conclusion:
The small values of $\Theta K N$ and other meson $N
\Theta$
couplings which such a small width implies, fall short
of explaining
the observed cross section for $\Theta$ production in
photon
initiated reactions. This naturally happens if the
$\Theta \rightarrow KN$ (or $K(*)N$, etc.) transitions
were
suppressed by a selection rule. The following
alternatives come to mind:
\begin{itemize}
\item (a) A new--hitherto unknown--quantum number is
possessed by $\Theta$(1540).

\item (b) The $\Theta$ is an I=2 isotensor.

\item (c) There are no strict new selection rules, but
the
complex ``topology" of the ``Color Network" in the
$\Theta$ penta-quark reduces its coupling to states with only
``simple" baryons and mesons.
\end{itemize}

Alternative (a) is most radical. Since the $\Theta$ is
produced together with ordinary non-exotic hadrons it
is difficult
to envision a new conserved quantum number. QCD,
flavor dynamics
and symmetries are well understood and radical, new
physics may be contemplated only if all other efforts
to explain the peculiarities of $\Theta(1540)$ fail.

The suggestive idea (b) \cite{PCR}  that isospin is
violated by a $K N \Theta$
(or any $K(*)-N-\Theta$)  immediately explains the
second puzzle pointed above. Unfortunately it is untenable:
First, the I=2 state should be much higher, \cite{N}
and, second, the other
members of the isospin quadruplet are missing.

We will focus here on alternative (c) which was
briefly alluded to before \cite{GN}. The idea is that
the new {\it narrow} exotic states are ground states in a
new family of hadrons. Ordinary $\bar{qq}$
mesons contain--in a chromoelectric flux
tube picture--just one flux tube connecting the $q$
and $\bar{q}$ and $qqq$ baryons have ``Y" shaped color
networks with the three flux tubes emanating from
the three quarks merging into one
``junction". The new qadri- and penta-quark states
consist of more complex networks with
junction--anti-junction
or two junctions--one anti-junction as indicated in
Fig. 1(a) and 1(b) of Ref.\cite{GN}.

Generic $\bar{qqq}\bar{q}$ or $\bar{qqqqq}$
meson-meson
or baryon-meson do not belong in the new family.
In collisions of ordinary hadrons transient
association due to hadron-hadron attraction
can form and may have some four or five quark,``single
bag" components. Such states are likely to have short
lifetimes of order $1/c$ with 1 $\sim$ Fermi hadronic
sizes and large O(200 MeV) widths. Their density
increases rapidly due
to coupling to multi-particle channels and these
{\it broad} exotic resonances then merge into a
continuum.

The longest range, one- or two-pion exchange, hadronic
interactions are attractive. For mesons containing
heavy $c/b$ quarks, such forces can suffice to form weakly bound
states \cite{T,MW}. It is important to distinguish two types of
quadri-quarks; namely,
$\bar{MM'} \{= \bar{Qq} \bar{Q'}q'\}$ and
$MM' \{= \bar{QQ'q}\bar{q'}\}$ states. The newly
discovered
$\bar{c}c \{(\bar{uu} + \bar{dd})/2^{(1/2)}\}$ in $B$
decays at Belle \cite{Belle} is of the first type.
Its small decay width  $\sim 20 \; MeV$ to $J/\psi+\pi \pi$ may
be due to a small overlap with the physically small
$\bar{c}c$ state.  To the extent that it can be viewed as,
say, $D(*)\bar{D}$ bound state, it also may not belong
in the new family considered.

The second type of $MM'$, say,
$DD(*)\{\bar{ccu}\bar{d}\}$  bound
states--if existing--are more likely to be of the
special
form of interest \cite{GeN}  here. The two heavy
quarks
$\bar{QQ'}$ and, separately, the two light
$\bar{q}\bar{q'}$ should
couple to a $\bar{3}\{3\}$ of color. Such couplings
like in
baryons, involve $\epsilon(abe)Q^a Q'^b$ and
$\epsilon(cde) \bar{q}(c)\bar{q'}(d)$, respectively,
and these
resulting structures with $\bar{3}$ and 3 SU(3) color,
should then join via $3^e \cdot \bar{3}_e$ to
make the overall color singlet state.

In the chromoelectric flux model this is represented
via a
juction/anti-junction where the two flux tubes
emerging
from $QQ'$ entering into $\bar{q}\bar{q'}$ are
incident. The
junctions in turn are connected by the same standard
``minimal flux"
tube going from the anti-junction to the junction.
Such $>-<$ coupling patterns occur also for systems
with light quarks
only, say,
$q^1 q^2 \bar{q}^3 \bar{q}^4$. We assume that the
$\Theta(1540) = \bar{s}(ud)^1(ud)^2$ has two
junctions, $J^1$ and $J^2$,
where the flux tubes of quarks in $(ud)^1$ and those
in $(ud)^2$
converge, respectively, and one anti-junction
$\bar{J}$
from which the three flux tubes ending at $\bar{s}$
and $J^1$ and $J^2$ emerge.
A key observation is that a quadri/penta-quark with
these color
networks can decay into two mesons/meson and baryon,
only if (a) junction and the anti-junction annihilate.
Also a $K$ or $K(*)$, etc., nucleon collision produce
the penta-quark
only if an extra  $J$ and $\bar{J}$ are ``pair"
created in
addition to the junction in the initial nucleon.

If such junction pair creation and annihilations are
suppressed, both difficulties pointed above may be
resolved. The decay width to the (only open) $KN$ channel
will be small and $\theta$
production via collision with a nucleon of real
or off-shell $K$'s, $K(*)$'s, $K(**)$'s, etc., may be
so small that an altogether different production
mechanism needs to be invoked.

Amusingly, $J-\bar{J}$ production was implicitly
discussed
in a paper on $q-\bar{q}$ pair production in the
chromoelectric
field inside the flux tube. \cite{CNN}

For any instantaneous, say, Red-$\bar{{\rm Red}}$
configuration
of the end $Q$ and $\bar{Q}$ anti-quarks and
corresponding $E$ fields,
production of a $\bar{{\rm red}}-{\rm red} \bar{q}q$
new pair is
preferred with the $\bar{{\rm red}}/{\rm red}
\bar{q}/q$ pulled
towards the Red/$\bar{{\rm Red}}$ end-quarks. For
$Q-\bar{Q}$ jets
in $e(+) e(-)$ colliders
this basic process repeats many times. It occurs
also when the energy available is more limited, say,
in the decay of an excited $\bar{Q}Q$ vector meson
produced in
$e(+) e(-)$ collision into two lighter $\bar{Q}q +
\bar{q}Q$ mesons
with $\bar{qq}$ creation occurring only once.

Denoting the field strength operative here by $E$, the
masses of the light quarks produced by $m$ and their
momentum
transverse to the $Q-\bar{Q}$ separation (or in high
energies,
the``jet" axis) by
$p$, the rate of $\bar{q}q$ pair creation is
proportional to:
\begin{equation}
  d(n)/d(p^2) \mid {\rm ``standard"} \sim \{(gE)^2\}
   exp - \{\pi \cdot [m^2+p^2]/(gE)\}
\label{dndpStandardEq12}
\end{equation}

Even in the above ``Red" field inside the flux we
have (due to peculiarities of SU(3) color) in addition
to the
preferred $\bar{r}r \bar{q}q$ pair production, {\it
also}
the production of blue-$\bar{{\rm blue}}$ or
white-$\bar{{\rm white}}$
light quarks. Unlike before, here the produced
quark/antiquark  is
attracted by and moves towards the
end-quark/antiquark. If the process
stops here, then a diquark/anti-diquark $Qq$ and
$\bar{q}\bar{Q}$ connected by a standard flux tube,
namely, the tetra-quark of interest is created!

Baryon--anti-baryon production happens
 when the missing white-$\bar{{\rm white}}$ (or
blue-$\bar{{\rm
blue}}$) pair is  created.)

The chromoelectric field strength relevant for this
``disfavored" production mode is $E/2$ rather than
$E$, yielding :
\begin{equation}
d(n)/d(p^2)\mid {\rm ``disfavored"} \,,\;\;{\rm
SU(3) \; color}
  \sim \{(g[E/2])^2\} exp-2  \{\pi [m^2+p^2]/(gE)\}
\label{dndpDisfavoredE/2Eq13}
\end{equation}
The  factor 1/2 $\{1/(N-1) \; {\rm for} \; SU(N)\}$
is readily explained: In the fundamental
representation 3 of $SU(3)$
drawn in two $\{= N-1 = {\rm rank \; of} \; SU(N)\}$
dimensions
the R, B, W quarks point along the directions of the
three complex
roots of unit:
$1,-1/2+i(3/4)^{(1/2)},-1/2-i(3/4)^{(1/2)}$. The
chromoelectric field of size $E$ produced by $R$ has a
component
-1/2 $E$ along B (or W) and a blue (or white)
quark--rather than
$\bar{r}$--can be produced but with half the effective
field strength.

Similarly the SU(N) fundamental representation is a
symmetric $N-1$ simplex and the angle between  any
pair of unit
vectors is $cos^{(-1)}[1/N-1]$. The analog of
Eq. (\ref{dndpDisfavoredE/2Eq13}) is then:
\begin{equation}
 d(n)/d(p^2) \mid {\rm ``disfavored"} \, \;\; {\rm
SU(N) \; of \;  color}
\sim \{g [E/(N-1)]\}^2 exp-(N-1)  \{\pi
[m^2+p^2]/(gE)\}
\label{dndpDisfavoredE/N-1Eq14}
\end{equation}
Note that for large $N(c)$ implicit in Skyrmion models,
the ``disfavored" mode is  exponentially suppressed in
$N$ which, indeed, is likely for baryon--anti-baryon and
monopole--anti-monopole pair production \cite{W,DN}
with $N \rightarrow 1/\alpha$.
(A similar exponential suppression is expected also in
the time-reversed process of annihilation of an N-fold
junction and anti-junction.  The suppression
can be understood in this case also in simple combinatorial
terms: each of the tubes of N colors in the junction
has to match up with the anti-tube of the same color in the
anti-junction. Thus only one out of 1/N! pairing
can lead to $\bar{J}J$ annihilation.)
For the $N=3$ case of interest we have the 1/4 in the
$E^2$ prefactor and
an extra suppression by $\sim$ .2 due to the doubled
exponent \cite{CNN}. The overall suppression
(1/10)-(1/20)
is consistent with the multiplicity of anti-nucleons
observed in jets or Z decays.

In gamma-nucleon collisions studied in the
above-mentioned
experiments the photon can virtually transform into a
pair of energetic $\bar{Q}\bar{Q}$ quarks $[Q=s]$ and the
diquarks next
form via the disfavored $\bar{u}u$ or $\bar{d}d$
creation as above.
Alternatively, the photon can impart a large energy to
one, say, $u$-quark
in the target nucleon and the disfavored
creation--now of $\bar{s}s$--happens later somewhere
along the resulting
prolonged flux tube originating at the struck
$u$-quark.

Once the ``junction barrier" has been overcome and an
intermediate entity like an $su >-< \bar{s}\bar{d}$
tetra-quark
has formed, the remaining process, Tetra+$N$
$\rightarrow$
Penta+$K$(or $K(*)$),  required to obtain the final
state of the
above experiments is straightforward. It involves only
standard quark exchanges and fusion/cutting of flux
tubes which are
familiar from ordinary meson-baryon and meson-meson
processes. Thus the above factor of (1/10)-(1/20)
approximates
the suppression of $\Theta$ production relative to
ordinary resonance in the above photonic experiments.
This concurs with the above estimates of the
``effective $\Theta$ width" of 5-15 MeV $\sim$
(1/10)-(1/20)
of normal hadronic widths.
Recall that the true width of $\Theta$ inferred
from independent purely hadronic $K$-neutron data is
smaller, say, O(1 MeV).
This small $\Gamma$ and
the disagreement with the effective width required in
``naive" $K$, $K(*)$, etc., exchange models constitute
the difficulties (1) {\it and} (2) above.

It has been argued \cite{GN} that longevity of some
Tetra-quarks and Penta-quark states may reflect the
difficulty of annihilating a junction $J$ and anti-junction
$\bar{J}$. This could be due to the smallness of the junction
radius $b$ as compared with the hadronic size  $\sim$ 0.7 Fermi.
The suppression becomes more dramatic
$\sim (b/a)^5$ if we have a centrifugal
barrier due to a relative $\ell=1$ angular momenta
between the junctions (or diquarks).
Such barriers are present when an isolated $\Theta$
decays into or forms out of a meson $K$ and baryon
$N$, but not
necessarily in the higher energy $\gamma-N$
collisions. This may  explain the apparent discrepancy
between the true
$\Gamma <$  0(1 MeV) and the effective $\Gamma$ of
5-15 MeV
required to explain the production rate.

The extra color dynamics-related suppression that the
early work implies for $J-\bar{J}$ production is
likely to affect also in the $J-\bar{J}$ annihilation
in tetra- and penta-decays. A future, more complete model
incorporating both this
with the earlier geometric size arguments for the
small width will hopefully provide a more compelling
explanation for the remarkably small $\Gamma(\Theta)$.

\noindent {\large \bf IV. Possible Manifestation of
the New Narrow Resonances
in Nucleon/Anti-Nucleon Annihilations and Some
Concluding Remarks}

We sketched in the previous section a possible
scenario for the anomalously small $\gamma(\theta)$ and the
apparent contradiction between the latter and $\Theta$
production rates in
photon-induced reactions which seem to require larger
widths. Is this scenario  viable?

One difficulty is the lack of evidence for narrow
tetra-quark states which our scenario requires.
The lightest member of this family
should not be heavier than $\sim$ 1200-1100 MeV--lying
300-400 MeV below
the $\theta$ penta-quark. (The scalar $a,f(980)$ may
indeed be four
quark/single bag states \cite{Jaffe1}. Yet the normal
widths of these
states  suggest that these are not the $>-<$ tetra $a$
that we discuss here.)

The formation of junction--anti-junction pairs or the
disappearance of
such, is the essence of $\bar{N}N$ pair
creation/annihilation. The
latter does {\it not} require that (anti-)quarks from
the respective (anti-)nucleon annihilate.  Rather,
$[\bar{q}\bar{q}\bar{q}]-qqq$ rearrange into
three $\bar{q}q$ pairs. These could be pseudoscalar,
vector and some higher  mesons.
For $\bar{p}p$ at rest the rate of ``genuine"
annihilations of $\bar{q}-q$'s is expected to be
larger than at higher energies. Annihilations of just
{\it one}
$\bar{q}q$ pair yield final states with {\it two},
rather than
three, mesons, happen in $\sim$ 25\% of the cases.

In the chromoelectric flux picture we can readily
envision  a
$q-\bar{q}$ annihilation occurring prior to the
annihilation of the
junction $J$ in the nucleon and the anti-junction
$\bar{J}$ in the anti-nucleon. Such events
involving the fusion of the
flux tube segments emerging from/terminating on the
specific $q$ and $\bar{q}$ which annihilated yield
tetra-quarks with
two junctions of the type considered here:
$>-q + \bar{q}-< \rightarrow  >--< $, namely, the
tetra-quark of interest.

One would expect to see in careful studies of
$p-\bar{p}$
annihilations at low energies in experiments like LEAR
these narrow resonances precisely in the "two meson" final
states.  Indeed, annihilation models favor formation of such
states.

The $N-\bar{N}$ potential  is attractive at all ranges
causing the initial $\bar{N}N$ to
accelerate and move towards each other. Thus,
annihilation at low energies
has a much larger cross section than the small
junction area $\pi \cdot b^2$ as
expected at high energies \cite{EG.N}.

The annihilation separates into two stages:
during the acceleration pions are emitted and
eventually the
$J-\bar{J}$ annihilate with further pions emitted.

Tetra-quarks can form at the end of
the first stage. If further we have $\ell=1$ between
the two junctions
the $(b/a)^5$ suppression of the $J \bar{J}$ \cite{GN}
may be operative and the tetra state can be narrow. Note,
however, that excited tetra
states decay to lower tetras via fast pionic
emissions. Only the
ground tetra state and very nearby higher states will
be narrow. If this state is as low as 1100-1200 MeV, then 0(3)
pions are emitted both prior to its formation and
in its decay.  The large combinatorial factor may
explain the absence of
these narrow tetra-quarks in the LEAR experiments
which focused on near $\bar{N}N$
threshold states recoiling against one photon or pion.
States which are within less than $m(\pi)$ from the
lowest tetra-quark
will decay emitting fairly sharp $\gamma$s of energy
$E \sim M(ex)-m(gr)$.

Full QCD lattice simulations recently performed for
baryons with three quarks  pinned down at relative
distances of O.7
Fermi clearly indicate via contours of equal action
density the ``Y" configuration with a narrow $b = 0.2$ Fermi
junction. \cite{JAP}  If this is so,  in reality then
the $b/a$ ratio of$\sim 0.3$ may lead to
a $(b/a)^5 \sim .25 \cdot 10^{-3}$ suppression of
$P$-wave quadri- and penta-quark
decays which is clearly sufficient for our purpose.
However, in ground state nucleon or mesons the fast
light quarks are
likely to tangle up the short flux tubes into a
uniform spherical distribution.
Note that spherical symmetry is not the issue: The
latter obtains for
S-wave   meson ground states, even if we had
``needle-like" narrow flux
tubes, by superposing, with equal amplitudes all
states $\mid \theta\phi >$
where the ``needle" points in a particular direction
on the unit sphere. Still, just to achieve semiclassical
constructs and narrow flux tubes,
in particular, we need to employ many quantum states
with high quantum
numbers. Can QCD generate such a rich family of states already at
energies of $\sim$ 1 GeV
in order to explain the peculiarities of these recent
experiments?  The same question can be rephrased as:
Is it conceivable that the complex $\Theta^{(+)}(1543)$
state that we envision with three junctions and seven
flux tube segments is that light?  In view of the title
of the paper, we surely hope that the answer is
in the affirmative.

\noindent {\bf Acknowledgements}

We are grateful to Ralf Gothe, Fred Muirer and Dave
Tedeschi for
helpful discussions and suggestions.

[Note added: After completing this work,
the paper by Liu and Ko \cite{LK} has been brought to
our attention. These careful calculations of $K K(*)$
exchanges confirm one
specific point which we made: namely, our estimated
large $\Theta$ width
required to explain the Saphir production cross
sections.]

\end{document}